\documentclass[preprint,superscriptaddress,aps,floatfix]{revtex4}
\usepackage{amsfonts}
\usepackage{graphics}
\usepackage{graphicx}
\usepackage{slashed}
\usepackage[utf8]{inputenc}
\newcommand{\be}{\begin{equation}}
\newcommand{\ee}{\end{equation}}
\newcommand{\en}{\end{equation}}
\newcommand{\ba}{\begin{eqnarray}}
\newcommand{\ea}{\end{eqnarray}}
\newcommand{\bea}{\begin{eqnarray}}
\newcommand{\eea}{\end{eqnarray}}

\providecommand*{\zeta}{\textqoppa}
\providecommand*{\zeta}{\textQoppa}

\begin{document}

\title{On one-loop corrections to the CPT-even Lorentz-breaking extension of QED}

\author{T. Mariz}
\affiliation{Instituto de F\'\i sica, Universidade Federal de Alagoas, 57072-270, Macei\'o, Alagoas, Brazil}
\email{tmariz@fis.ufal.br}

\author{R. V. Maluf}
\email{r.v.maluf@fisica.ufc.br}
\affiliation{Departamento de F\'{\i}sica, Universidade Federal do Cear\'{a},\\
 Caixa Postal 6030, 60455-760, Fortaleza, CE, Brazil}

\author{J. R. Nascimento}
\email{jroberto@fisica.ufpb.br}
\affiliation{Departamento de F\'{\i}sica, Universidade Federal da Para\'{\i}ba\\
 Caixa Postal 5008, 58051-970, Jo\~ao Pessoa, Para\'{\i}ba, Brazil}

\author{A. Yu. Petrov}
\email{petrov@fisica.ufpb.br}
\affiliation{Departamento de F\'{\i}sica, Universidade Federal da Para\'{\i}ba\\
 Caixa Postal 5008, 58051-970, Jo\~ao Pessoa, Para\'{\i}ba, Brazil}

\begin{abstract}

In this paper, we describe the quantum electrodynamics added by Lorentz-violating CPT-even terms in the context of the standard model extension. We focus our attention on the fermion sector, represented by the CPT-even symmetric Lorentz-breaking tensor $c_{\mu\nu}$. We adopt a generic form that parametrizes the components of $c_{\mu\nu}$ in terms of one four-vector, namely, $c_{\mu\nu}=u_\mu u_\nu - \zeta \frac{u^2}{4}g_{\mu\nu}$. We then generate perturbatively, up to the third order in this tensor, the aether-like term for the gauge field. Finally, we discuss the renormalization scheme for the gauge propagator, by taking into account $c_{\mu\nu}$ traceless ($\zeta=1)$ and, trivially, $c_{\mu\nu}=u_\mu u_\nu$ ($\zeta=0$).
\end{abstract}

\maketitle

\section{Introduction}

The possible violation of the Lorentz symmetry is intensively discussed now. A general framework that describes the CPT and Lorentz symmetry breaking at low-energy level is the so-called the Standard-Model Extension (SME). A typical manner of its introduction is based on an extension of the corresponding action by new terms involving constant vector or tensor fields which explicitly break the Lorentz symmetry by introducing the privileged direction of the space-time. Many examples of such additive terms are presented in \cite{Kostel,Kostel1,Kostel2}. This matter naturally calls the interest to study of different issues related to such models, both at the classical and at the quantum levels. Certainly, extensions of gauge field theories are of special interest. It was shown that in the tree approximations, such theories display highly nontrivial effects such as birefringence of waves and rotation of the plane of polarization in the vacuum (see f.e. \cite{Ja,MP}).

In the quantum level, these theories exhibit even more interesting properties. The paradigmatic example is the QED with the Carroll-Field-Jackiw (CFJ) term \cite{CFJ} which is known to be gauge invariant, and, being generated as a quantum correction, it is finite but ambiguous (an  incomplete list of references on the CFJ term is given by \cite{list}). These studies certainly call the interest to consideration of quantum properties of other Lorentz-breaking extensions of the QED. In particular, the studies of the CPT-even Lorentz-breaking extensions of the QED are of particular importance. At the classical level, many issues related to such extensions, especially exact solutions and dispersion relations, were studied in \cite{aethercl}, and the paper \cite{Carroll} opened an interest in these theories from the viewpoint of the extra dimension concept. Further, in \cite{aether}, the CPT-even terms were shown to arise as quantum corrections in a CPT-odd extended QED with a nonminimal coupling. Moreover, they turn out to be finite.

While the theory considered in \cite{aether} is, first, CPT-odd, second, involving non-renormalizable couplings (other interesting studies on quantum corrections in non-renormalizable Lorentz-breaking extensions of QED are presented in \cite{start}), the natural question is -- whether the CPT-even contributions can be generated from an essentially CPT-even Lorentz-breaking extension of the QED? We note that the natural way to distinguish between CPT-even corrections arising either from CPT-odd or from CPT-even sectors is the following. The lower possible CPT-even correction can be already of the first order in the CPT-even Lorentz-breaking parameter, as we will show in this paper, but it must be of the second order  in the CPT-odd Lorentz-breaking parameter, see \cite{aether}, therefore, it is natural to expect that the contribution from the CPT-even sector will dominate since the Lorentz-breaking parameters are very small.
In this paper, we use CPT-even terms in the Maxwell and fermion sectors of SME, represented by the tensor $c_{\mu\nu}$, introduced in Ref. \cite{Carroll} and those ones in which the Lorentz-violating coefficient is traceless, with the resulting quadratic action being essentially renormalizable in the four-dimensional space-time. 

Some preliminary results on renormalization of this model, including the lower-order renormalization constants, were obtained already in \cite{Kostel,Kostel1}. Other important results on renormalization of Lorentz-breaking theories can be found in \cite{Altschul,Cambiaso,Scarp},  see also references therein. An alternative approach to the CPT-even Lorentz-breaking extension of QED, involving interesting geometrical analogies, is also presented in \cite{Drummond}. However, it is certainly interesting to obtain the next order aether-like counterterms. In fact, we will see that, for our two cases, the usual renormalization constant of gauge propagator can be used to renormalize the two contributions to the modified Maxwell actions obtained, respectively.

The structure of the paper looks like follows. In section 2, we introduce the renormalizable CPT-even extension of the QED. In section 3, we calculate the one-loop contributions to the two-point function of the gauge field, up to the third order in $c_{\mu\nu}$.  Finally, in the Summary, we discuss our results.

\section{The model}

We start with the following extended QED with a CPT-even Lorentz-breaking term (see f.e. \cite{Kostel}), given by
\begin{equation}
S=\int d^{4}x\left\{\bar{\psi}\left[i({g}^{\mu\nu}+c^{\mu\nu})\gamma_{\mu}\left(\partial_{\nu}+ieA_{\nu}\right)-m\right]\psi -\frac{1}{4}F_{\mu\nu}F^{\mu\nu}-\frac14(k_F)_{\mu\nu\lambda\rho}F^{\mu\nu}F^{\lambda\rho}\right\},\label{eq:action1}
\end{equation}
with $c_{\mu\nu}=u_\mu u_\nu - \zeta \frac{u^2}{4}g_{\mu\nu}$ and $g^{\mu\nu}=\mbox{diag}(1,-1,-1,-1)$,  where $c_{\mu\nu}$ is traceless (for $\zeta=1$) and, trivially, $c_{\mu\nu}=u_\mu u_\nu$ (for $\zeta=0$). We can ensure the smallness of $c_{\mu\nu}$ requiring that $|c_{\mu\nu}|\ll 1$ for any $\mu,\nu$. The coefficient  $(k_F)_{\mu\nu\lambda\rho}$, which can be written in terms of $c_{\mu\nu}$, i.e., as $(k_F)_{\mu\nu\lambda\rho}=g_{\mu\lambda}c_{\nu\rho}+g_{\nu\rho}c_{\mu\lambda}-g_{\mu\rho}c_{\nu\lambda}-g_{\nu\lambda}c_{\mu\rho}$, is double traceless, ${(k_F)_{\mu\nu}}^{\mu\nu}=0$, for $\zeta=1$, and ${(k_F)_{\mu\nu}}^{\mu\nu}\neq0$, for $\zeta=0$.
It is worth mentioning that there are other CPT-even terms in the minimal SME, i.e., those ones involving $d_{\mu\nu}$ and $H_{\mu\nu}$ (see f.e. \cite{Kostel1}). These terms will not be taken into account in this work since quantum corrections do not mix them with the $c$ and $k_{F}$ Lorentz-violating coefficients. 

Our goal is to find the contributions to the purely gauge sector in the one-loop order. So, let us write the purely spinor part of the action (\ref{eq:action1}), which we will use for quantum calculations:
\begin{equation}
S_\psi=\int d^{4}x\bar{\psi}\left(i\slashed{\partial}-m+ic^{\mu\nu}\gamma_{\mu}\partial_{\nu}-e\slashed{A}-ec^{\mu\nu}\gamma_{\mu}A_{\nu}\right)\psi.
\end{equation}
The essential feature of this theory is its renormalizability in four dimensions. Indeed, all constants in the theory are dimensionless.

Now, let us derive the Feynman rules for this theory.
The kinetic term for the spinor field is
\begin{equation}
S_{kin}=\int d^{4}x\bar{\psi}\left(i\slashed{\partial}-m+ic^{\mu\nu}\gamma_{\mu}\partial_{\nu}\right)\psi.
\end{equation}
Then, the corresponding propagator is
\begin{equation}\label{prop}
\left\langle \psi(p)\bar{\psi}(-p)\right\rangle \equiv i G(p)=\frac{i}{\slashed{p}-m+c^{\mu\nu}\gamma_{\mu}p_{\nu}}.
\end{equation}
Within this paper, we will adopt the perturbative expansion for the free propagator:
\begin{eqnarray}
\frac{i}{\slashed{p}-m+c^{\mu\nu}\gamma_{\mu}p_{\nu}} & \simeq & \frac{i}{\slashed{p}-m}\nonumber \\
 & + & \frac{i}{\slashed{p}-m}(ic^{\mu\nu}\gamma_{\mu}p_{\nu})\frac{i}{\slashed{p}-m}.\label{eq:FreeProp1}
\end{eqnarray}
The interaction term
\begin{eqnarray}
S_{int} & = & \int d^{4}x\bar{\psi}\left(-e\slashed{A}-ec^{\mu\nu}\gamma_{\mu}A_{\nu}\right)\psi
\end{eqnarray}
gives rise to the vertices
\begin{eqnarray}
V_1&=&-e\bar{\psi}\gamma^{\mu}\psi A_\mu,\\
V_2&=&-e\bar{\psi}\gamma_\nu c^{\mu\nu}\psi A_\mu.
\end{eqnarray}

\begin{figure}[ht]
 \centering
   \centerline{\includegraphics{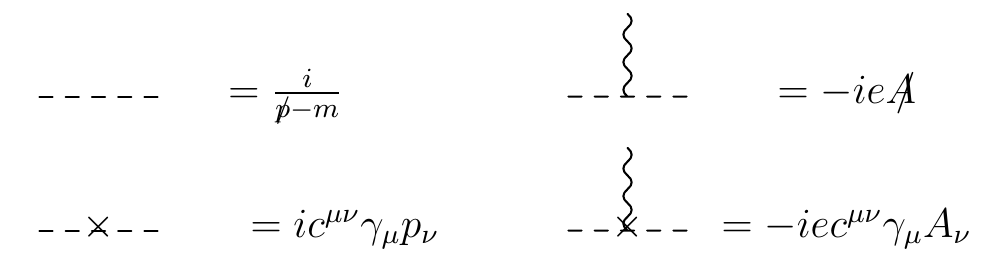}}
 \caption{Feynman rules.\label{fig1}}
 \end{figure}

Graphically, these Feynman rules can be represented by the Fig. \ref{fig1}. They will be used to introduce one-loop Feynman diagrams.

\section{Radiative corrections and induced terms}

The fermionic determinant in our theory can be read off from the generating functional
\begin{equation}
Z[A_\mu] = \int D\bar\psi D\psi e^{i\int d^4x{\cal L}_\psi} = e^{iS_{eff}},
\end{equation}
so that, by integrating over the fermions, we obtain the one-loop effective action 
\begin{equation}\label{Seff}
S_{eff} = -i\mathrm{Tr}\ln(\slashed{p}-m+\slashed{c}\cdot p-e\slashed{A}-e\slashed{c}\cdot A).
\end{equation}
Here, $\mathrm{Tr}$ stands for the trace over the Dirac matrices, as well as the trace over the integration in momentum and coordinate spaces, and, for brevity, we have introduced the notation $\slashed{c}\cdot p=\slashed{c}_\mu p^\mu$, with $\slashed{c}_\mu=c_{\mu\nu}\gamma^\nu$.

In order to single out the quadratic terms in $A_\mu$ of the effective action,  we initially rewrite the expression (\ref{Seff}) as
\begin{equation}
S_{eff}=S_{eff}^{(0)}+\sum_{n=1}^\infty S_{eff}^{(n)},
\end{equation}
where $S_{eff}^{(0)}=-i\mathrm{Tr}\ln(\slashed{p}-m+\slashed{c}\cdot p)$ and 
\begin{equation}
S_{eff}^{(n)} = \frac{i}{n}\mathrm{Tr}\left[\frac{1}{\slashed{p}-m+\slashed{c}\cdot p}(e\slashed{A}+e\slashed{c}\cdot A)\right]^n.
\end{equation}
Then, after evaluating the trace over the coordinate space, by using the commutation relation $A_\mu(x) G(p)=G(p-i\partial)A_\mu(x)$ and the completeness relation of the momentum space, for the quadratic action $S_{eff}^{(2)}$, we get
\begin{equation}\label{Seff2}
S_{eff}^{(2)} = \frac{1}{2}\int d^4x\, \Pi^{\mu\nu} A_\mu A_\nu,
\end{equation}
where
\begin{equation}\label{Pi}
\Pi^{\mu\nu} = ie^2\int\frac{d^{4}p}{(2\pi)^4}\mathrm{tr}\,G(p)\gamma^\mu G(p-i\partial)\gamma^\nu,
\end{equation}
with
\begin{equation}\label{Gb}
G(p) = \frac{1}{\slashed{p}-m+\slashed{c}\cdot p}
\end{equation}
being the Feynman propagator  (see Eq.~(\ref{prop})). Note that the derivative contained in $\Pi^{\mu\nu}$ acts only on the first gauge field $A_\mu$, in Eq.~(\ref{Seff2}).

For the zero order in $c_{\mu\nu}$, in momentum space, we have
\begin{equation}
\Pi_{0}^{\mu\nu} = ie^2\mathrm{tr} \int\frac{d^{4}p}{(2\pi)^4}S(p)\gamma^\mu S(p-k)\gamma^\nu,
\end{equation}
with $S(p)=(\slashed{p}-m)^{-1}$, which yields a paradigmatic result of the usual QED presented in textbooks, see f.e. \cite{Fuji}:
\begin{equation}
\Pi_{0}^{\mu\nu} = \left(\frac{e^2}{6 \pi ^2 \epsilon'}+A_0(\eta)\right)\left(k^{\mu} k^{\nu}-k^2 g^{\mu\nu}\right),
\end{equation}
where $\frac{1}{\epsilon'} = \frac{1}{\epsilon}-\ln\frac{m}{\mu'}$, with $\epsilon=4-D$ and $\mu'^2=4\pi\mu^2e^{-\gamma-i\pi}$, and
\begin{equation}
A_0(\eta) = \frac{e^2}{36 \pi ^2 k^4}\left[k^4 \left(6\eta\tan^{-1}\eta+5\right)-12 k^2 m^2 \left(\eta\tan^{-1}\eta-1\right)-48m^4\eta\tan^{-1}\eta\right],
\end{equation}
with $\eta^2=\frac{k^2}{4m^2-k^2}$. Then, for small $k^2$, we have the well-known result
\begin{equation}\label{pi0}
\Pi_{0}^{\mu\nu} = \left(\frac{e^2}{6 \pi ^2 \epsilon'}+\frac{e^2k^2}{60\pi^2m^2}\right)\left(k^{\mu} k^{\nu}-k^2 g^{\mu\nu}\right) + \cdots.
\end{equation}
Then, for the effective action, we obtain the usual pole part
\begin{equation}\label{Seff20}
S_{eff}^{(2,0)} = -\frac{e^2}{6 \pi ^2 \epsilon}\int d^4x\, \frac{1}{4}F_{\mu\nu}F^{\mu\nu}.
\end{equation}

\begin{figure}[ht]
\centerline{\includegraphics{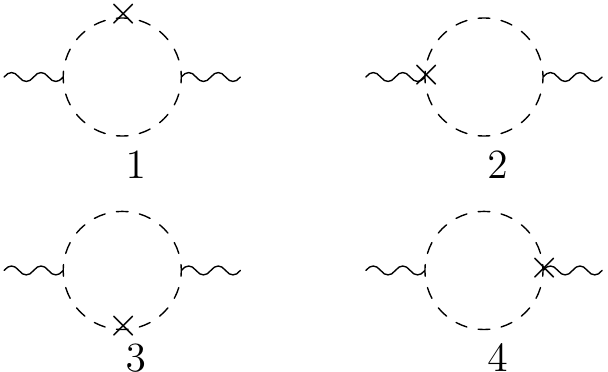}} 
\caption{First-order Lorentz-breaking contributions.\label{fig2}}
\end{figure}

From the graphical viewpoint, the first-order Lorentz-breaking corrections are depicted in Fig. \ref{fig2}.

\noindent Explicitly, the contribution of the first order in $c_{\mu\nu}$ is $\Pi_{1}^{\mu\nu}=\Pi_{1,1}^{\mu\nu}+\Pi_{1,2}^{\mu\nu}+\Pi_{1,3}^{\mu\nu}+\Pi_{1,4}^{\mu\nu}$ (see also \cite{Kostel}), where
\begin{equation}
\Pi_{1,1}^{\mu\nu} = -ie^2\mathrm{tr} \int\frac{d^{4}p}{(2\pi)^4}S(p)\slashed{c}\cdot pS(p)\gamma^\mu S(p-k) \gamma^\nu,
\end{equation}
\begin{equation}
\Pi_{1,2}^{\mu\nu} = -ie^2\mathrm{tr} \int\frac{d^{4}p}{(2\pi)^4}S(p)\gamma^\mu S(p-k)\slashed{c}\cdot (p-k)S(p-k) \gamma^\nu,
\end{equation}
\begin{equation}
\Pi_{1,3}^{\mu\nu} = ie^2\mathrm{tr} \int\frac{d^{4}p}{(2\pi)^4}S(p)\slashed{c}^\mu S(p-k)\gamma^\nu,
\end{equation}
\begin{equation}
\Pi_{1,4}^{\mu\nu} = ie^2\mathrm{tr} \int\frac{d^{4}p}{(2\pi)^4}S(p)\gamma^\mu S(p-k)\slashed{c}^\nu.
\end{equation}
After the calculation of the trace and the integrals, the total result is
\begin{eqnarray}
\label{pic}
\Pi_{1}^{\mu\nu} &=& \frac{e^2}{6 \pi ^2 \epsilon'}\left[\left(u^2 k^2-2 (u\cdot k)^2\right) g^{\mu\nu}+2 u^{\mu} \left(k^{\nu} (u\cdot k)-k^2 u^{\nu}\right)+k^{\mu} \left(2 u^{\nu} (u\cdot k)-u^2 k^{\nu}\right)\right] \nonumber\\
&&+ A_1(\eta)u^2(k^2 g^{\mu\nu}-k^\mu k^\nu)+ B_1(\eta) (u\cdot k)^2 g^{\mu\nu}+ C_1(\eta) \left[u^{\mu } \left(k^{\nu } (u\cdot k)-k^2 u^{\nu }\right) \right. \nonumber\\
&&\left.+k^{\mu } u^{\nu } (u\cdot k)\right] + D_1(\eta) (u\cdot k)^2 k^\mu k^\nu,
\end{eqnarray}
with
\begin{eqnarray}
A_1(\eta) &=& \frac{e^2\eta}{36 \pi ^2 k^4}\left[k^4 \left(6\tan^{-1}{\eta}-5 \eta \right)+k^2 m^2 \left(\eta  (9 \zeta+8)-12\tan^{-1}{\eta}\right)\right. \nonumber\\
&&\left.-12 m^4 (3 \zeta-4) \left(\eta -\tan^{-1}{\eta}\right)\right],
\end{eqnarray}
\begin{equation}
B_1(\eta) = \frac{e^2\eta}{9 \pi ^2 k^4}\left[k^4 \left(\eta -3\tan^{-1}\eta\right)+k^2 m^2 \left(6\tan^{-1}\eta-7 \eta \right)+12 m^4 \left(\eta -\tan^{-1}\eta\right)\right],
\end{equation}
\begin{equation}
C_1(\eta) =\frac{e^2}{18 \pi ^2 k^2 \eta }\left[k^2 \left(5 \eta -6\tan^{-1}\eta\right)+12 m^2 \left(\eta -\tan^{-1}\eta\right)\right],
\end{equation}
\begin{equation}
D_1(\eta) = \frac{e^2\eta}{6 \pi ^2 k^6} \left[k^2 \eta  \left(k^2+2 m^2\right)+24 m^4 \left(\tan^{-1}\eta-\eta \right)\right].
\end{equation}
Note that the finite contributions depend on $\zeta$, whereas the divergent ones do not. We can easily observe that for small $k^2$, we obtain 
\begin{eqnarray}
\label{pic1}
\Pi_{1}^{\mu\nu} &=& \frac{e^2}{6 \pi ^2 \epsilon'}\left[\left(u^2 k^2-2 (u\cdot k)^2\right) g^{\mu\nu}+2 u^{\mu} \left(k^{\nu} (u\cdot k)-k^2 u^{\nu}\right)+k^{\mu} \left(2 u^{\nu} (u\cdot k)-u^2 k^{\nu}\right)\right] \nonumber\\
&&-\frac{e^2u^2\zeta}{24\pi^2}\left(k^{\mu} k^{\nu}-k^2 g^{\mu\nu}\right) + \frac{e^2}{120 \pi ^2 m^2}\left\{k^{\mu } \left[4 k^2 u^{\nu } u\cdot k+k^{\nu } \left(4 (u\cdot k)^2-(\zeta+2)u^2 k^2 \right)\right]\right. \nonumber\\
&&\left.+k^2 \left[\left((\zeta+2)u^2 k^2 -8 (u\cdot k)^2\right)g^{\mu  \nu }+4 u^{\mu } \left(k^{\nu } (u\cdot k)-k^2 u^{\nu }\right)\right]\right\}+\cdots.
\end{eqnarray}

One can check that the pole part of this expression matches the known result \cite{Kostel}.
It is clear that this self-energy tensor is transversal. Manifestly, the corresponding divergent contribution to the effective action is
\begin{equation}\label{Seff21}
S_{eff}^{(2,1)} = \frac{e^2}{6 \pi ^2 \epsilon} \int d^4x\left(\frac{u^2}{4}F_{\mu\nu}F^{\mu\nu}-u^{\mu}F_{\mu\nu}u_{\lambda}F^{\lambda\nu}\right),
\end{equation}
which replays the structures of Maxwell term and the aether term \cite{Carroll}.
It should be observed that if one splits the constant tensor $u_{\mu}u_{\nu}$ into a sum of  traceless and trace parts, i.e.,
$u_{\mu}u_{\nu}=\bar c_{\mu\nu}+\frac{1}{4}u^2g_{\mu\nu}$ (with $\zeta=1$), respectively, this divergent contribution will be rewritten as
\begin{equation}\label{Seff21tl}
S_{eff}^{(2,1)} = -\frac{e^2}{6 \pi ^2 \epsilon}\int d^4x\, \bar c^{\mu}_{\:\:\lambda} F_{\mu\nu}F^{\lambda\nu},
\end{equation}
i.e., it is completely expressed in terms of the traceless $c_{\mu\nu}$, namely, $\bar c_{\mu\nu}$.

\begin{figure}[ht]
\centerline{\includegraphics{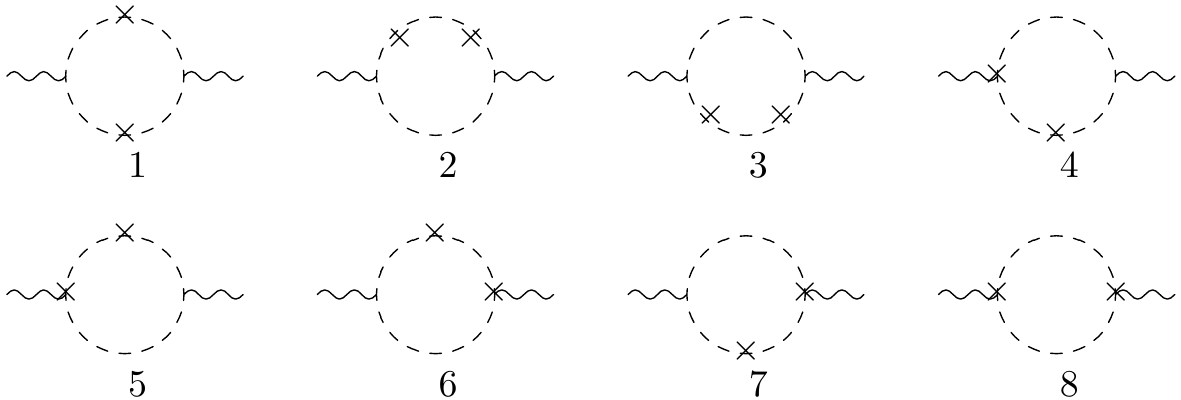}} 
\caption{Second-order Lorentz-breaking corrections.\label{fig3}}
\end{figure}

Now, let us perform the next step which naturally consists in calculating of the second-order aether-like quantum corrections which never considered earlier. For the second order in $c_{\mu\nu}$, we have $\Pi_{2}^{\mu\nu}=\Pi_{2,1}^{\mu\nu}+\Pi_{2,2}^{\mu\nu}+\Pi_{2,3}^{\mu\nu}+\Pi_{2,4}^{\mu\nu}+\Pi_{2,5}^{\mu\nu}+\Pi_{2,6}^{\mu\nu}+\Pi_{2,7}^{\mu\nu}+\Pi_{2,8}^{\mu\nu}$, with
\begin{equation}
\Pi_{2,1}^{\mu\nu} = ie^2\mathrm{tr} \int\frac{d^{4}p}{(2\pi)^4}S(p)\slashed{c}\cdot pS(p)\slashed{c}\cdot pS(p)\gamma^\mu S(p-k) \gamma^\nu,
\end{equation}
\begin{equation}
\Pi_{2,2}^{\mu\nu} = ie^2\mathrm{tr} \int\frac{d^{4}p}{(2\pi)^4}S(p)\slashed{c}\cdot pS(p)\gamma^\mu S(p-k)\slashed{c}\cdot (p-k)S(p-k) \gamma^\nu,
\end{equation}
\begin{equation}
\Pi_{2,3}^{\mu\nu} = ie^2\mathrm{tr} \int\frac{d^{4}p}{(2\pi)^4}S(p)\gamma^\mu S(p-k)\slashed{c}\cdot (p-k)S(p-k)\slashed{c}\cdot (p-k)S(p-k) \gamma^\nu,
\end{equation}
\begin{equation}
\Pi_{2,4}^{\mu\nu} = -ie^2\mathrm{tr} \int\frac{d^{4}p}{(2\pi)^4}S(p)\slashed{c}\cdot pS(p)\slashed{c}^\mu S(p-k) \gamma^\nu,
\end{equation}
\begin{equation}
\Pi_{2,5}^{\mu\nu} = -ie^2\mathrm{tr} \int\frac{d^{4}p}{(2\pi)^4}S(p)\slashed{c}^\mu S(p-k)\slashed{c}\cdot (p-k)S(p-k) \gamma^\nu,
\end{equation}
\begin{equation}
\Pi_{2,6}^{\mu\nu} = -ie^2\mathrm{tr} \int\frac{d^{4}p}{(2\pi)^4}S(p)\slashed{c}\cdot pS(p)\gamma^\mu S(p-k)\slashed{c}^\nu,
\end{equation}
\begin{equation}
\Pi_{2,7}^{\mu\nu} = -ie^2\mathrm{tr} \int\frac{d^{4}p}{(2\pi)^4}S(p)\gamma^\mu S(p-k)\slashed{c}\cdot (p-k)S(p-k)\slashed{c}^\nu,
\end{equation}
\begin{equation}
\Pi_{2,8}^{\mu\nu} = ie^2\mathrm{tr} \int\frac{d^{4}p}{(2\pi)^4}S(p)\slashed{c}^\mu S(p-k)\slashed{c}^\nu.
\end{equation}
From the graphical viewpoint, the second-order Lorentz-breaking corrections are depicted in Fig. \ref{fig3}.
Then, the result is
\begin{eqnarray}
\label{picc0}
\Pi_{2}^{\mu\nu} &=& \frac{e^2 u^2}{6 \pi^2\epsilon'}\left[\left((u\cdot k)^2-u^2 k^2\right) g^{\mu\nu}+u^{\mu} \left(k^2 u^{\nu}-k^{\nu} (u\cdot k)\right)+k^{\mu} \left(u^2 k^{\nu}-u^{\nu} (u\cdot k)\right)\right] \nonumber\\
&&+A_2(\eta) u^2 u^{\mu } u^{\nu }+B_2(\eta) u^4 k^2 g^{\mu  \nu }+C_2(\eta) u^4 k^{\mu } k^{\nu } +D_2(\eta) u^2 (k^{\mu }u^{\nu }+u^{\mu }k^{\nu }) (u\cdot k)  \nonumber\\
&&+ E_2(\eta) u^2 (u\cdot k)^2 g^{\mu  \nu }+ F_2(\eta)u^2 (u\cdot k)^2 k^{\mu } k^{\nu }+ G_2(\eta) (u\cdot k)^2u^{\mu } u^{\nu }\nonumber\\
&&+ H_2(\eta) (k^{\mu }u^{\nu }+u^{\mu } k^{\nu }) (u\cdot k)^3+I_2(\eta)(u\cdot k)^4 g^{\mu  \nu } + J_2(\eta) (u\cdot k)^4k^{\mu } k^{\nu }.
\end{eqnarray}
In Appendix, we present the explicit expressions for the factors $A_2(\eta)\ldots J_2(\eta)$, for brevity. However, for small $k^2$, we can write 
\begin{eqnarray}
\label{picc}
\Pi_{2}^{\mu\nu} &=&\frac{e^2 u^2}{24 \pi^2\epsilon'}\left[g^{\mu  \nu } \left(u^2 k^2 (\zeta-4)-2 (\zeta-2) (u\cdot k)^2\right)+2 (\zeta-2) u^{\mu } \left(k^{\nu } (u\cdot k)-k^2 u^{\nu }\right)\right. \nonumber\\
&&\left.+k^{\mu } \left(2 (\zeta-2) u^{\nu } (u\cdot k)-u^2 (\zeta-4) k^{\nu }\right)\right] + \frac{e^2u^2 \zeta}{192 \pi ^2} \left[g^{\mu  \nu } \left(u^2 k^2 (\zeta-6) \right.\right.\nonumber\\
&&\left.\left.+12 (u\cdot k)^2\right)+12 u^{\mu } \left(k^2 u^{\nu }-k^{\nu } (u\cdot k)\right)-k^{\mu } \left(12 u^{\nu } (u\cdot k)+u^2 (\zeta-6) k^{\nu }\right)\right] \nonumber\\
&&-\frac{e^2}{960 \pi ^2 m^2}\left\{g^{\mu  \nu } \left[u^4 k^4 ((\zeta+4) \zeta+16)-16 u^2 k^2 (\zeta+2) (u\cdot k)^2+64 (u\cdot k)^4\right] \right.\nonumber\\
&&-8 u^{\mu } \left(k^{\nu } (u\cdot k)-k^2 u^{\nu }\right) \left(8 (u\cdot k)^2-u^2 k^2 (\zeta+2)\right)+k^{\mu } \left[u^2 k^{\nu } \left(8 (\zeta+2) (u\cdot k)^2\right.\right. \nonumber\\
&&\left.\left.\left.-u^2 k^2 ((\zeta+4) \zeta+16)\right)+8 u^{\nu } (u\cdot k) \left(u^2 k^2 (\zeta+2)-8 (u\cdot k)^2\right)\right]\right\} + \cdots.
\end{eqnarray}

It is interesting to note that, although this self-energy tensor is also transversal, it differs from (\ref{pic1}), since the Maxwell term and the aether term enter this contribution with weights different from (\ref{pic1}).
The purely divergent contribution to the effective action from this sector is
\begin{equation}\label{Seff22}
S_{eff}^{(2,2)} = \frac{e^2}{6 \pi ^2 \epsilon}\int d^4x \left(\frac{u^4}{16}(\zeta-4)F_{\mu\nu}F^{\mu\nu}-\frac{u^2}{4}(\zeta-2)u^{\mu}F_{\mu\nu}u_{\lambda}F^{\lambda\nu}\right).
\end{equation}
So, we succeeded to find aether-like one-loop divergences. We note that unlike the first-order contributions, the second-order ones essentially involve both $u^2$ and the traceless $\bar{c}_{\mu\nu}$. The expression for the effective action then becomes 
\bea\label{Seff22tl}
S_{eff}^{(2,2)} = \frac{e^2}{6 \pi ^2 \epsilon} \int d^4x \left(-\frac{u^4}{8}F_{\mu\nu}F^{\mu\nu}+\frac{u^2}{4}\bar c^{\mu}_{\:\:\lambda}F_{\mu\nu}F^{\lambda\nu}\right),
\eea
where we have used again $u_{\mu}u_{\nu}=\bar c_{\mu\nu}+\frac{1}{4}u^2g_{\mu\nu}$, with $\zeta=1$, in Eq.~(\ref{Seff22}).

Let us finally consider the third-order Lorentz-breaking  quantum correction, in order to try to obtain a general expression for these divergent contributions. Thus, for the third order in $c_{\mu\nu}$, we must calculate $\Pi_{3}^{\mu\nu}=\Pi_{3,1}^{\mu\nu}+\Pi_{3,2}^{\mu\nu}+\Pi_{3,3}^{\mu\nu}+\Pi_{3,4}^{\mu\nu}+\Pi_{3,5}^{\mu\nu}+\Pi_{3,6}^{\mu\nu}+\Pi_{3,7}^{\mu\nu}+\Pi_{3,8}^{\mu\nu}+\Pi_{3,9}^{\mu\nu}+\Pi_{3,10}^{\mu\nu}+\Pi_{3,11}^{\mu\nu}+\Pi_{3,12}^{\mu\nu}$, where
\begin{equation}
\Pi_{3,1}^{\mu\nu} = -ie^2\mathrm{tr} \int\frac{d^{4}p}{(2\pi)^4}S(p)\slashed{c}\cdot pS(p)\slashed{c}\cdot pS(p)\slashed{c}\cdot pS(p)\gamma^\mu S(p-k) \gamma^\nu,
\end{equation}
\begin{equation}
\Pi_{3,2}^{\mu\nu} = -ie^2\mathrm{tr} \int\frac{d^{4}p}{(2\pi)^4}S(p)\slashed{c}\cdot pS(p)\slashed{c}\cdot pS(p)\gamma^\mu S(p-k)\slashed{c}\cdot (p-k)S(p-k) \gamma^\nu,
\end{equation}
\begin{equation}
\Pi_{3,3}^{\mu\nu} = -ie^2\mathrm{tr} \int\frac{d^{4}p}{(2\pi)^4}S(p)\slashed{c}\cdot pS(p)\gamma^\mu S(p-k)\slashed{c}\cdot (p-k)S(p-k)\slashed{c}\cdot (p-k)S(p-k) \gamma^\nu,
\end{equation}
\begin{equation}
\Pi_{3,4}^{\mu\nu} = -ie^2\mathrm{tr} \int\frac{d^{4}p}{(2\pi)^4}S(p)\gamma^\mu S(p-k)\slashed{c}\cdot (p-k)S(p-k)\slashed{c}\cdot (p-k)S(p-k)\slashed{c}\cdot (p-k)S(p-k) \gamma^\nu,
\end{equation}
\begin{equation}
\Pi_{3,5}^{\mu\nu} = ie^2\mathrm{tr} \int\frac{d^{4}p}{(2\pi)^4}S(p)\slashed{c}\cdot pS(p)\slashed{c}\cdot pS(p)\slashed{c}^\mu S(p-k) \gamma^\nu,
\end{equation}
\begin{equation}
\Pi_{3,6}^{\mu\nu} = ie^2\mathrm{tr} \int\frac{d^{4}p}{(2\pi)^4}S(p)\slashed{c}\cdot pS(p)\slashed{c}^\mu S(p-k)\slashed{c}\cdot (p-k)S(p-k) \gamma^\nu,
\end{equation}
\begin{equation}
\Pi_{3,7}^{\mu\nu} = ie^2\mathrm{tr} \int\frac{d^{4}p}{(2\pi)^4}S(p)\slashed{c}^\mu S(p-k)\slashed{c}\cdot (p-k)S(p-k)\slashed{c}\cdot (p-k)S(p-k) \gamma^\nu,
\end{equation}
\begin{equation}
\Pi_{3,8}^{\mu\nu} = ie^2\mathrm{tr} \int\frac{d^{4}p}{(2\pi)^4}S(p)\slashed{c}\cdot pS(p)\slashed{c}\cdot pS(p)\gamma^\mu S(p-k) \slashed{c}^\nu,
\end{equation}
\begin{equation}
\Pi_{3,9}^{\mu\nu} = ie^2\mathrm{tr} \int\frac{d^{4}p}{(2\pi)^4}S(p)\slashed{c}\cdot pS(p)\gamma^\mu S(p-k)\slashed{c}\cdot (p-k)S(p-k) \slashed{c}^\nu,
\end{equation}
\begin{equation}
\Pi_{3,10}^{\mu\nu} = ie^2\mathrm{tr} \int\frac{d^{4}p}{(2\pi)^4}S(p)\gamma^\mu S(p-k)\slashed{c}\cdot (p-k)S(p-k)\slashed{c}\cdot (p-k)S(p-k) \slashed{c}^\nu,
\end{equation}
\begin{equation}
\Pi_{3,11}^{\mu\nu} = -ie^2\mathrm{tr} \int\frac{d^{4}p}{(2\pi)^4}S(p)\slashed{c}\cdot pS(p)\slashed{c}^\mu S(p-k)\slashed{c}^\nu.
\end{equation}
\begin{equation}
\Pi_{3,12}^{\mu\nu} = -ie^2\mathrm{tr} \int\frac{d^{4}p}{(2\pi)^4}S(p)\slashed{c}^\mu S(p-k)\slashed{c}\cdot (p-k)S(p-k)\slashed{c}^\nu.
\end{equation}

\begin{figure}[ht]
\centerline{\includegraphics{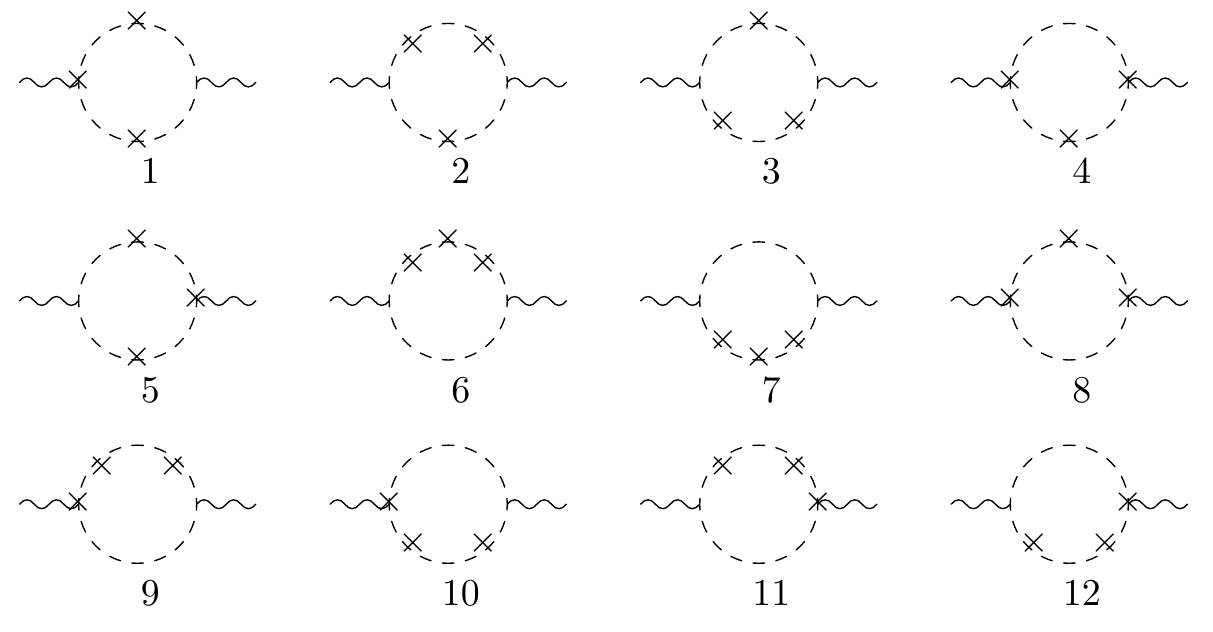}} 
\caption{Third-order Lorentz-breaking corrections.\label{fig4}}
\end{figure}

These third-order corrections are depicted in Fig. \ref{fig4}. 

Then, the total result can be written as
\begin{eqnarray}\label{Pi3}
\Pi_{3}^{\mu\nu} &=& \frac{e^2 u^4}{96 \pi^2\epsilon'} \left(g^{\mu  \nu } \left(u^2 k^2 (\zeta-4)^2-2 ((\zeta-4) \zeta+8) (u\cdot k)^2\right) +2 ((\zeta-4) \zeta+8) \right. \nonumber\\
&&\left.\times u^{\mu } \left(k^{\nu } (u\cdot k)-k^2 u^{\nu }\right)+k^{\mu } \left(2 ((\zeta-4) \zeta+8) u^{\nu } (u\cdot k)-u^2 (\zeta-4)^2 k^{\nu }\right)\right) \nonumber\\
&&+ \text{finite terms},
\end{eqnarray}
where we have calculated only the divergent terms. We have omitted the finite terms because up to now we are interested in interring a general expression for the renormalization constant of the gauge propagator. Thus, by taking into account (\ref{Seff2}), the effective action becomes
\begin{equation}\label{Seff23}
S_{eff}^{(2,3)} = \frac{e^2}{6 \pi ^2 \epsilon}\int d^4x\left(\frac{u^6}{64}(\zeta-4)^2F_{\mu\nu}F^{\mu\nu}-\frac{u^4}{16}((\zeta-4)\zeta+8)u^{\mu}F_{\mu\nu}u_{\lambda}F^{\lambda\nu}\right).
\end{equation}
Rewriting the above expression in terms of traceless $\bar{c}_{\mu\nu}$, we obtain
\begin{equation}\label{Seff23tl}
S_{eff}^{(2,3)} = \frac{e^2}{6 \pi ^2 \epsilon}\int d^4x \left(\frac{u^6}{16}F_{\mu\nu}F^{\mu\nu}-\frac{5}{16}u^4\bar c^{\mu}_{\:\:\lambda} F_{\mu\nu}F^{\lambda\nu}\right).
\end{equation}

Finally, the complete one-loop divergent contribution to the two-point function, for $\zeta=0$, is given by the sum of (\ref{Seff20},\ref{Seff21},\ref{Seff22},\ref{Seff23}). Thus, we have
\begin{equation}
S_{eff}^{(2)} = -\frac{e^2}{6 \pi ^2 \epsilon}\int d^4x \left(\frac{1}{4}(1-u^2+u^4-u^6)F_{\mu\nu}F^{\mu\nu}+\left(1-\frac{u^2}{2}+\frac{u^4}{2}\right)u^{\mu}F_{\mu\nu}u_{\lambda}F^{\lambda\nu}\right) + \cdots.
\end{equation}
By analyzing the above expression, we can easily write the general pole part result 
\begin{equation}
S_{eff}^{(2)} = -\frac{e^2}{6 \pi ^2 \epsilon}\int d^4x\, \frac{1}{4}\left(1+u^2\right)^{-1}\tilde F_{\mu\nu} \tilde F^{\mu\nu},
\end{equation}
where
\begin{equation}
\tilde F_{\mu\nu} = (g_{\mu\alpha}+u_\mu u_\alpha)(g_{\nu\beta}+u_\nu u_\beta) F^{\alpha\beta},
\end{equation}
which is, in fact, the expression we should have when consider the nonperturbative propagator (\ref{prop}).  Formally, here we have an effective metric $\tilde{g}_{\mu\alpha}=g_{\mu\alpha}+u_\mu u_\alpha$, and $S_{eff}^{(2)}$ is proportional to $\tilde{g}_{\mu\nu}\tilde{g}_{\alpha\beta}F^{\mu\alpha}F^{\nu\beta}$, which replays the form of the Lagrangian of the electromagnetic field in a curved space. Thus, we must modify the Lorentz-violating Maxwell action in (\ref{eq:action1}), as follows:
\begin{equation}\label{SA1}
S_{A}= -\frac14 \int d^4x\, \left(1+u^2\right)^{-1} \tilde F_{\mu\nu} \tilde F^{\mu\nu},
\end{equation}
so that the renormalization constant is given by the usual one
\begin{equation}\label{Z3}
Z_3 = 1 - \frac{e^2}{6 \pi ^2 \epsilon},
\end{equation}
in which the coefficient $\left(1+u^2\right)^{-1}$ can be absorbed in the renormalization constant of the generating functional. Note that we can rewrite the modified Maxwell Lagrangian as
\begin{equation}
{\cal L}_A = -\frac14 \tilde F_{\mu\nu} \tilde F^{\mu\nu} = -\frac14 F_{\mu\nu} F^{\mu\nu} -\frac14(k_F)_{\mu\nu\lambda\rho} F^{\mu\nu}F^{\lambda\rho},
\end{equation}
where
\begin{equation}\label{kF1}
(k_F)_{\mu\nu\lambda\rho} = \left(1+\frac{u^2}{2}\right) (g_{\mu\lambda}u_\nu u_\rho+g_{\nu\rho}u_\mu u_\lambda-g_{\mu\rho}u_\nu u_\lambda-g_{\nu\lambda}u_\mu u_\rho).
\end{equation}
This expression for $(k_F)_{\mu\nu\lambda\rho}$  was used in \cite{Gomes:2009wn}, in order to keep the Lagrangian formally covariant, as we are observing, however, here, it is the first time that it has been obtained through radiative corrections. 

Now, let us discuss the more interesting case, when $c_{\mu\nu}$ is traceless, by considering the sum of  (\ref{Seff20},\ref{Seff21tl},\ref{Seff22tl},\ref{Seff23tl}). The effective action of the pole part is given by
\begin{equation}
S_{eff}^{(2)} = -\frac{e^2}{6 \pi ^2 \epsilon}\int d^4x \left(\frac{1}{4}\left(1+\frac{u^4}{2}-\frac{u^6}{4}\right)F_{\mu\nu}F^{\mu\nu}+\left(1-\frac{u^2}{4}+\frac{5u^4}{16}\right)\bar c^{\mu}_{\:\:\lambda} F_{\mu\nu}F^{\lambda\nu}\right) + \cdots.
\end{equation}
Then, by analyzing the above equation, the general result is trivially written as
\begin{equation}
S_{eff}^{(2)} = -\frac{e^2}{6 \pi ^2 \epsilon}\int d^4x\, \frac{1}{4}\left(1+\frac{3u^2}{4}\right)^{-1}\left(1-\frac{u^2}{4}\right)^{-3}\bar F_{\mu\nu} \bar F^{\mu\nu},
\end{equation}
with now
\begin{equation}
\bar F_{\mu\nu} = (g_{\mu\alpha}+\bar c_{\mu\alpha})(g_{\nu\beta}+\bar c_{\nu\beta}) F^{\alpha\beta},
\end{equation}
so, one has the effective metric $\bar g_{\mu\alpha}=g_{\mu\alpha}+\bar c_{\mu\alpha}$ which again allows to write the $S_{eff}^{(2)}$ as a result proportional to $\bar g_{\mu\nu}\bar g_{\alpha\beta}F^{\mu\alpha}F^{\nu\beta}$.
Therefore, for this case, the modified Maxwell action must assume the form
\begin{equation}\label{SA2}
S_{A}= -\frac14 \int d^4x\, \left(1+\frac{3u^2}{4}\right)^{-1}\left(1-\frac{u^2}{4}\right)^{-3} \bar F_{\mu\nu} \bar F^{\mu\nu},
\end{equation}
where again the renormalization constant is the usual one (\ref{Z3}). By rewriting the above action in terms of traceless $\bar{c}_{\mu\nu}$, we also obtain the Lagrangian 
\begin{equation}
{\cal L}_A = -\frac14 \bar F_{\mu\nu} \bar F^{\mu\nu} = -\frac14 F_{\mu\nu} F^{\mu\nu} -\frac14(k_F)_{\mu\nu\lambda\rho} F^{\mu\nu}F^{\lambda\rho},
\end{equation}
with
\begin{equation}\label{kF2}
(k_F)_{\mu\nu\lambda\rho} =(4 g_{\mu\lambda}\bar c_{\nu\rho}+4 \bar c_{\mu\lambda}\bar c_{\nu\rho}+2g_{\mu\lambda}\bar c_{\nu\alpha}\bar c^{\alpha}_{\:\:\rho}+4\bar c_{\mu\lambda}\bar c_{\nu\alpha}\bar c^{\alpha}_{\:\:\rho}+\bar c_{\mu\alpha}\bar c^{\alpha}_{\:\:\lambda} \bar c_{\nu\beta}\bar c^{\beta}_{\:\:\rho}),
\end{equation}
which is more involved than (\ref{kF1}), but can be reduced to it by replacing, formally, $\bar c_{\mu\nu}$  by $u_\mu u_\nu$, i.e., we get $(k_F)_{\mu\nu\lambda\rho} =(4+2u^2) g_{\mu\lambda}u_\nu u_\rho$.

Let us now discuss the general structure of quantum corrections of different orders in insertions. The examples presented above showed that the generic result with $n$ insertions will look like
\begin{equation}
S_{eff,\, n}^{(2)} = -\frac{e^2}{\pi ^2 \epsilon}\int d^4x \left(a_nu^{2n}F_{\mu\nu}F^{\mu\nu}+b_nu^{2n-2}\bar c^{\mu}_{\:\:\lambda} F_{\mu\nu}F^{\lambda\nu}\right),
\end{equation}
where the coefficients $a_n$, $b_n$ are some numbers. It is the only possible form for such corrections. However, one cannot determinate a precise form of the $a_n$, $b_n$ without explicit calculations, and there is no any {\it a priori} reason for existence of any relations between these coefficients.

\section{Summary}

In this paper, for the CPT-even Lorentz-breaking extension of QED, we showed that the aether term naturally arises as a quantum correction, and succeeded to calculate the unique renormalized modified Maxwell action to the third order in a constant symmetric second-rank tensor $c_{\mu\nu}$, for which, in its part, we considered all possible structures, from the traceless one up to the direct product of two constant vectors. In the four-dimensional case, it diverges which requires an introduction of corresponding counterterms, both in Maxwell and in aether sectors. Actually, it involves only one renormalization constant (\ref{Z3}), so that, in principle, one needs to modify the Maxwell action already at the tree level (see Eqs.~(\ref{SA1}) and (\ref{SA2})). By rewriting these actions in terms of $(k_F)_{\mu\nu\lambda\rho}$, we obtained the expressions (\ref{kF1}) and (\ref{kF2}), respectively, which are the all-insertion contributions for the CPT-even Lorentz-breaking extension of Maxwell theory. It is interesting to note that for the light-like $u_{\mu}$, only the lower order in $u_{\mu}$ yields nontrivial contributions both to the Maxwell and to the aether terms. We note that, unlike \cite{Drummond}, in the four-dimensional space-time we succeeded to calculate not only the divergent part but also the finite one of the CPT-even term.  Also, the results which should be mentioned here are, first, the fact that within our calculations we for the first time obtained explicitly through radiative corrections the covariantized Lagrangian (\ref{kF1}) which earlier \cite{Gomes:2009wn} was shown to arise within a completely distinct context, that is, the calculation of the free energy; second, we generalized this coefficient for the essentially traceless $c_{\mu\nu}$; third, we demonstrated that the result, involving both Maxwell and aether terms, is characterized by only one renormalization constant.

The importance of our results consists in the fact that in the four-dimensional space-time our theory is renormalizable in all orders in all $c_{\mu\nu}$ insertions, just as the electrodynamics with the CFJ term \cite{CFJ}. Another important feature of our theory consists in arising of an effective metric looking like $\tilde{g}_{\mu\nu}=g_{\mu\nu}+u_{\mu}u_{\nu}$ or $\bar{g}_{\mu\nu}=g_{\mu\nu}+\bar c_{\mu\nu}$. In principle, the calculations in resulting theory obtained through summation of the tree-level Maxwell term and the quantum correction can be carried out with use of this effective metric implying in appropriate modification of vertices and propagators formulated in the new space with the metric $\tilde{g}_{\mu\nu}$ or $\bar{g}_{\mu\nu}$. We note that in \cite{Scarp}, where the effective metric arises due to the Lorentz-breaking modification of the Dirac matrices, the results were obtained, first, only in the spinor sector, second, only up to the first order in Lorentz-breaking parameters, and in \cite{Drummond} the effective metric has been introduced already on the step of consideration of the classical action, while we show how it arises as a result of summation of quantum corrections of different orders in $\bar c_{\mu\nu}$ or $u_{\mu}u_{\nu}$. Therefore the space-time turns out to be the affine one (see also discussions in \cite{Scarp,Drummond}), as it occurs for the CPT-even supersymmetric Lorentz-breaking theories \cite{CM}. It is natural to expect that it can open the way to implement the Lorentz symmetry breaking within the supergravity. The possible way to do it is as follows: we consider the Lorentz-breaking supersymmetry algebra  discussed in \cite{CM}, and suggest the Lorentz-breaking tensor $c_{\mu\nu}$ (which in \cite{CM} has been chosen in the simplest form $c_{\mu\nu}=u_{\mu}u_{\nu}$) to be a fixed function of space-time coordinates rather than constant. This clearly would imply in a more sophisticated algebra of supercovariant derivatives, and in this case the effective metric clearly will not be more an affine one, implying in a non-zero curvature, hence we face a problem of studying the supersymmetric theory on a curved background, that is, coupled to a supergravity. We expect to study this problem in one of our next papers.

{\bf Acknowledgements.} This work was partially supported by Conselho
Nacional de Desenvolvimento Cient\'{\i}fico e Tecnol\'{o}gico (CNPq). The work by A. Yu. P. has been supported by the
CNPq project No. 303783/2015-0.

\vspace*{3mm}

\centerline{\bf APPENDIX}

In this Appendix, we present the explicit results for the factors $A_2(\eta)\ldots J_2(\eta)$ (see Eq.~(\ref{picc0})).
\begin{eqnarray}
A_2(\eta) &=& \frac{e^2\left(k^2-4 m^2\right)^{-1}}{288 \pi ^2 \eta  k^6}\left[k^{10} \left(\eta  \left(-3 \eta ^2-26 \zeta+37\right)+24\left(2 \eta ^2 (\zeta+1)+3 \zeta\right)\tan^{-1}\eta\right) \right.\nonumber\\
&&\left.+576 \eta ^2 k^2 m^8 (\zeta-4) \left(3 \eta ^2\tan^{-1}\eta+\eta +\tan^{-1}\eta\right)-48 \eta  k^4 m^6 \left(\eta ^2 (11 \zeta-34) \right.\right.\nonumber\\
&&\left.\left.+ \eta  \left(9 \eta ^2 (\zeta-4)+46 \zeta-56\right)\tan^{-1}\eta+3 (\zeta-4)\right)+12 k^6 m^4 \left(2 \eta  \left(\eta ^2 (\zeta-7) \right.\right.\right.\nonumber\\
&&\left.\left.\left.+9 (4 \zeta-3)\right)+3\left(\eta ^2 \left(\eta ^2 (\zeta-4)+33 \zeta-4\right)-8 \zeta\right)\tan^{-1}\eta\right)+2 k^8 m^2 \right.\nonumber\\
&&\left.\times\left(\eta  \left(\eta ^2 (9 \zeta-6)-35 \zeta-44\right)-36 \left(\eta ^2 (5 \zeta+4)+3 \zeta\right)\tan^{-1}\eta\right) \right.\nonumber\\
&&\left.-2304 \eta ^4 m^{10} (\zeta-4)\tan^{-1}\eta\right],
\end{eqnarray}
\begin{eqnarray}
B_2(\eta) &=& \frac{e^2\left(k^2-4 m^2\right)^{-1}}{1152 \pi ^2 \eta k^8}\left[k^{10} \left(\eta  \left(19 \eta ^2+52 \zeta-141\right)-48 \left(2 \eta ^2 (\zeta+2)+3 \zeta\right)\tan^{-1}\eta\right) \right. \nonumber\\
&&\left.-2304 \eta ^2 k^2 m^8 \left(\left(\eta ^2 (\zeta (4 \zeta+7)-12)+\zeta-4\right)\tan^{-1}\eta+\eta  (\zeta-4)\right) +96 \eta  k^4 m^6 \right.\nonumber\\
&&\left.\times\left(\eta ^2 \left(24 \zeta^2+34 \zeta-53\right)+2 \eta \left(3 \eta ^2 (\zeta (9 \zeta+11)-12)+26 \zeta-56\right)\tan^{-1}\eta \right.\right.\nonumber\\
&&\left.\left.+6 (\zeta-4)\right)-8 k^6 m^4 \left(\eta  \left(7 \eta ^2 (6 \zeta (3 \zeta+2)-11)+3 (92 \zeta-93)\right)+18 \left(\eta ^2 \left(\eta ^2 \right.\right.\right.\right.\nonumber\\
&&\left.\left.\left.\left.\times(5 \zeta (\zeta+1) -4)+13 \zeta
-4\right)-8 \zeta\right)\tan^{-1}\eta\right)+4 k^8 m^2 \left(\eta  \left(\eta ^2 \left(27 \zeta^2-14\right) \right.\right.\right.\nonumber\\
&&\left.\left.\left.+44 \zeta+69\right)+72 \left(2 \eta ^2 (\zeta+2)+\zeta\right)\tan^{-1}\eta\right) +9216 \eta ^4 m^{10} (\zeta-4)\tan^{-1}\eta\right],
\end{eqnarray}
\begin{eqnarray}
C_2(\eta) &=& \frac{e^2\left(k^2-4 m^2\right)^{-1}}{1440 \pi ^2 \eta  k^8}\left[k^{10} \left(\eta  \left(-26 \eta ^2-65 \zeta+174\right)+60\left(2 \eta ^2 (\zeta-1)+3 (\zeta-2)\right) \right.\right.\nonumber\\
&&\left.\left.\times\tan^{-1}\eta\right)+5760 \eta ^2 k^2 m^8 \left(\left(\eta ^2 (2 \zeta (\zeta+1)-3)-1\right)\tan^{-1}\eta-\eta \right)-240 \eta k^4 m^6 \right.\nonumber\\
&&\left.\times\left(2 \left(\eta ^2 (6 \zeta(\zeta+1)-7)-3\right)+\eta\left(3 \eta ^2 (\zeta (9 \zeta+8)-6)+44 (\zeta-1)\right)\tan^{-1}\eta\right) \right.\nonumber\\
&&\left.+4 \eta  k^6 m^4\left(\eta ^2 (45 \zeta(7 \zeta+4)-104)+45 \eta\left(\eta ^2 (\zeta (5 \zeta+4)-2)+28 \zeta-34\right)\tan^{-1}\eta \right.\right.\nonumber\\
&&\left.\left.+660 \zeta-600\right)-k^8 m^2 \left(\eta  \left(\eta ^2 \left(135 \zeta^2-88\right)+4 (55 \zeta+84)\right)+360 \left(\eta ^2 (3 \zeta-4) \right.\right.\right.\nonumber\\
&&\left.\left.\left.+2 (\zeta-2)\right)\tan^{-1}\eta\right)+23040 \eta^4 m^{10}\tan^{-1}\eta\right],
\end{eqnarray}
\begin{eqnarray}
D_2(\eta) &=& \frac{e^2\left(k^2-4 m^2\right)^{-1}}{1440 \pi ^2 \eta  k^8}\left[k^{10} \left(\eta  \left(11 \eta ^2+130 \zeta-189\right)-120 \left(\eta ^2 (2 \zeta-1)+3 (\zeta-1)\right) \right.\right.\nonumber\\
&&\left.\left.\times\tan^{-1}\eta\right)+2880 \eta ^2 k^2 m^8 \left(\left(\eta ^2 (17 \zeta+6)+6 \zeta+2\right)\tan^{-1}\eta+2 \eta  (3 \zeta+1)\right) \right.\nonumber\\
&&\left.-240 \eta  k^4 m^6 \left(\eta ^2 (39 \zeta+14)+\eta  \left(6 \eta ^2 (8 \zeta+3)-31 \zeta+44\right)\tan^{-1}\eta+18 \zeta +6\right) \right.\nonumber\\
&&\left.+4 \eta  k^6 m^4 \left(\eta ^2 (315 \zeta+134)+45 \eta \left(\eta ^2 (5 \zeta+2)-43 \zeta+34\right)\tan^{-1}\eta-645 \zeta \right.\right. \nonumber\\
&&\left.\left.+600\right)+2 k^8 m^2 \left(\eta  \left(-29 \eta ^2+220 \zeta+153\right)+360 \left(\eta ^2 (3 \zeta-2)+2 (\zeta-1)\right) \right.\right.\nonumber\\
&&\left.\left.\times\tan^{-1}\eta\right)-23040 \eta ^4 m^{10} (3 \zeta+1)\tan^{-1}\eta\right],
\end{eqnarray}
\begin{eqnarray}
E_2(\eta) &=& \frac{e^2\left(k^2-4 m^2\right)^{-1}}{720 \pi ^2 \eta  k^8}\left[k^2 \left(-k^8 \left(\eta  \left(49 \eta ^2+35 \zeta+9\right)-60\left(\eta ^2 (2 \zeta-7)+3 (\zeta-3)\right) \right.\right.\right.\nonumber\\
&&\left.\left.\left.\times\tan^{-1}\eta\right)-120 \eta  k^2 m^6 \left(5 \eta ^2 (88-3 \zeta)+ \eta\left(6 \eta ^2 (9 \zeta+95)-35 \zeta+394\right)\tan^{-1}\eta \right.\right.\right.\nonumber\\
&&\left.\left.\left.-36 \zeta+186\right)+2 \eta  k^4 m^4 \left(\eta ^2 (1035 \zeta+4258)+45 \eta\left(\eta ^2 (19 \zeta+64)+27 \zeta-16\right) \right.\right.\right.\nonumber\\
&&\left.\left.\left.\times\tan^{-1}\eta-165 \zeta+4050\right)-4 k^6 m^2 \left(\eta  \left(\eta ^2 (90 \zeta+7)+40 \zeta+126\right)+90   \left(\eta ^2  \right.\right.\right.\right.\nonumber\\
&&\left.\left.\left.\left.\times(3 \zeta-10)+2 (\zeta-3)\right)\tan^{-1}\eta\right)+1440 \eta ^2 m^8 \left(\left(\eta ^2 (188-13 \zeta)-12 \zeta+62\right) \right.\right.\right.\nonumber\\
&&\left.\left.\left.\times\tan^{-1}\eta+2 \eta  (31-6 \zeta)\right)\right)+11520 \eta^4 m^{10} (6 \zeta-31)\tan^{-1}\eta\right],
\end{eqnarray}
\begin{eqnarray}
F_2(\eta) &=& \frac{e^2}{240 \pi ^2 k^{10}}\left[k^8 \left(19 \eta ^2-10 \zeta+39\right)+480 \eta  k^2 m^6 \left(\left(\eta ^2 (6 \zeta+20)+9\right)\tan^{-1}\eta+9 \eta \right) \right.\nonumber\\
&&\left.-120 k^4 m^4 \left(2 \eta ^2 (3 \zeta+7)+ \eta\left(\eta ^2 (8 \zeta+11)-2 \zeta+15\right)\tan^{-1}\eta+9\right)+2 k^6 m^2 \right. \nonumber\\
&&\left.\times\left(\eta ^2 (60 \zeta+37)-30 \zeta+135\right)-17280 \eta ^3  m^8\tan^{-1}\eta\right],
\end{eqnarray}
\begin{eqnarray}
G_2(\eta) &=& \frac{e^2}{720 \pi ^2 k^8}\left[k^8 \left(11 \eta ^2+360 \left(\eta ^2+1\right)\eta^{-1}\tan^{-1}\eta+251\right)+1440 \eta  k^2 m^6 \left(\left(29 \eta ^2+16\right) \right.\right.\nonumber\\
&&\left.\left.\times\tan^{-1}\eta+16 \eta \right)-120 k^4 m^4 \left(\left(39 \eta ^3+87 \eta \right)\tan^{-1}\eta+55 \eta ^2+48\right)+2 k^6 m^2 \right.\nonumber\\
&&\left.\times\left(83 \eta ^2-720\eta\tan^{-1}\eta+825\right)-92160 \eta^3 m^8\tan^{-1}\eta\right],
\end{eqnarray}
\begin{eqnarray}
H_2(\eta) &=& \frac{e^2}{240 \pi ^2 k^{10}}\left[\left(\eta ^2-79\right) k^8-480 \eta  k^2 m^6 \left(\left(16 \eta ^2+9\right)\tan^{-1}\eta+9 \eta \right) +120 k^4 m^4 \left(10 \eta ^2 \right.\right.\nonumber\\
&&\left.\left.+\left(7 \eta ^2+23\right) \eta\tan^{-1}\eta+9\right)-34 \left(\eta ^2+15\right) k^6 m^2+17280 \eta ^3 m^8\tan^{-1}\eta\right],
\end{eqnarray}
\begin{eqnarray}
I_2(\eta) &=& \frac{e^2}{240 \pi ^2 k^{10}}\left[\left(29 \eta ^2+69\right) k^8+480 \eta  k^2 m^6 \left(\left(34 \eta ^2+9\right)\tan^{-1}\eta+9 \eta \right)-120 k^4 m^4  \right.\nonumber\\
&&\left.\times\left(28 \eta ^2+\left(29 \eta ^2+21\right)\eta\tan^{-1}\eta+9\right)+2 \left(167 \eta ^2+225\right) k^6 m^2-17280 \eta ^3 m^8 \right.\nonumber\\
&&\left.\tan^{-1}\eta\right],
\end{eqnarray}
\begin{eqnarray}
J_2(\eta) &=& -\frac{e^2\eta^2}{6 \pi ^2 k^{10}}\left[-12 k^2 m^4 \left(6 \eta\tan^{-1}\eta+5\right)+8 k^4 m^2+k^6+240 \eta m^6\tan^{-1}\eta\right].
\end{eqnarray}


\begin{thebibliography}{50}
\bibitem{Kostel} D. Colladay, V. A. Kostelecky, Phys. Rev. D {\bf 55}, 6770 (1997), hep-ph/9703464; Phys. Rev. D {\bf 58}, 116002 (1998), hep-ph/9809521. 
\bibitem{Kostel1}V. A. Kostelecky, C. D. Lane, A. G. M. Pickering, Phys.Rev. D65 (2002) 056006, hep-th/0111123. 
\bibitem{Kostel2} V. A. Kostelecky, Phys. Rev. D {\bf 69}, 105009 (2004), hep-th/0312310.
\bibitem{Ja} R. Jackiw, Nucl. Phys. Proc. Suppl. 108, 30 (2002), hep-th/0110057.
\bibitem{MP} R. Myers and M. Pospelov, Phys. Rev. Lett. 90, 211601 (2003), hep-ph/0301124.
\bibitem{CFJ} S. Carroll, G. Field, R. Jackiw, Phys. Rev. D {\bf 41}, 1231 (1990).
\bibitem{list} R. Jackiw, V. A. Kostelecky, Phys. Rev. Lett. 82, 3572 (1999), hep-ph/9901358;
S. Coleman and S. L. Glashow, Phys. Rev. D {\bf59}, 116008 (1999), hep-ph/9812418;
M. P\'erez-Victoria, Phys. Rev. Lett. {\bf83}, 2518 (1999), hep-th/9905061;
J. M. Chung and P. Oh, Phys. Rev. D {\bf60}, 067702 (1999), hep-th/9812132;
J. M. Chung, Phys. Rev. D {\bf60}, 127901 (1999), hep-th/9904037; Phys. Lett. B {\bf461}, 138 (1999), hep-th/9905095;
W. F. Chen, Phys. Rev. D {\bf60}, 085007 (1999), hep-th/9903258; J. M. Chung and B. K. Chung, 105015 (2001), hep-th/0101097; C.~Adam and F.~R.~Klinkhamer, Phys.\ Lett.\ B {\bf 513}, 245 (2001), hep-th/0105037.
\bibitem{aethercl} H. Belich, T. Costa-Soares, M. M. Ferreira Jr., J. A. Helayel-Neto, Eur. Phys. J. C41, 421 (2005), hep-th/0410104; Eur. Phys. J. C42, 127 (2005), hep-th/0411151; H. Belich, T. Costa-Soares, M. M. Ferreira Jr., J. A. Helayel-Neto, F. M. O. Moucherek, Phys. Rev. D74, 065009 (2006), hep-th/0604149; F. Klinkhamer, M. Schreck, Nucl. Phys. B848, 90 (2011), arXiv: 1011.4258; F. Klinkhamer, M. Schreck, Nucl. Phys.B856, 666 (2011), arXiv: 1110.4101; M. Schreck, Phys. Rev. D86, 065038 (2012), arXiv: 1111.4182; Phys. Rev. D89, 085013 (2014), arXiv: 1311.0032.
\bibitem{Carroll} S. Carroll, H. Tam, Phys. Rev. D78, 044047 (2008), arXiv: 0802.0521.
\bibitem{aether} M. Gomes, J. R. Nascimento, A. Yu. Petrov, A. J. da Silva, Phys. Rev. D81, 045018 (2010), arXiv: 0911.3548; "On the Aether-like Lorentz-breaking action for the electromagnetic field", arXiv: 1008.0607; G. Gazzola, H. G. Fargnoli, A. P. Baeta Scarpelli, M. Sampaio, M. C. Nemes, J. Phys. G39, 035002 (2012), arXiv: 1012.3291; A. P. Baeta Scarpelli, T. Mariz, J. R. Nascimento, A. Yu. Petrov, Eur. Phys. J. C73, 2526 (2013), arXiv: 1304.2256.
\bibitem{start} R. Casana, M. M. Ferreira, E. O. Silva, E. Passos, F. E. P.
dos Santos, Phys. Rev. D87, 047701 (2013), arXiv: 1212.6361; R. Casana, M. M. Ferreira, R. Maluf, F. E. P. dos Santos,
Phys. Lett. B726, 815 (2013), arXiv: 1302.2375.
\bibitem{Altschul} A. Ferrero, B. Altschul, Phys. Rev. D84, 065030 (2011), arXiv: 1104.4778.
\bibitem{Cambiaso} M. Cambiaso, R. Lehnert, R. Potting, Phys. Rev. D85, 085023 (2012), arXiv: 1201.3045.
\bibitem{Scarp} L. C. T. Brito, H. G. Fargnoli, A. P. Baeta Scarpelli, Phys. Rev. D87, 125023 (2013), arXiv: 1304.6016; M. Cambiaso, R. Lehnert, R. Potting, Phys.Rev. D90, 065003 (2014), arXiv: 1401.7317. 
\bibitem{Drummond}I.~T.~Drummond, Phys.\ Rev.\ D {\bf 95}, 025006 (2017), arXiv:1603.09211.
\bibitem{Fuji} K. Fujikawa, H. Suzuki, Path Integrals and Quantum Anomalies, Clarendon Press, Oxford, 2004.
\bibitem{Gomes:2009wn}M.~Gomes, T.~Mariz, J.~R.~Nascimento, A.~Y.~Petrov, A.~F.~Santos and A.~J.~da Silva, Phys.\ Rev.\ D {\bf 81}, 045013 (2010), arXiv:0910.4560.
\bibitem{CM} D. Colladay, P. McDonald, Phys. Rev. D83, 025021 (2011), arXiv: 1010.1781; A. C. Lehum, J. R. Nascimento, A. Yu. Petrov, A. J. da Silva, Phys. Rev. D88, 045022 (2013), arXiv: 1305.1812. 

\end{thebibliography}
\end{document}